\documentclass[preprint,aps,showpacs]{revtex4}
\usepackage[dvips]{graphicx}

\begin{document}

\newcommand{\dfrac}[2]{\frac{\displaystyle #1}{\displaystyle #2}}
\preprint{VPI--IPPAP--01--06}

\title{The Effect of the Minimal Length Uncertainty Relation on
the Density of States and the Cosmological Constant Problem}
\author{
Lay~Nam~Chang\footnote{electronic address: laynam@vt.edu},
Djordje~Minic\footnote{electronic address: dminic@vt.edu},
Naotoshi~Okamura\footnote{electronic address: nokamura@vt.edu}, and
Tatsu~Takeuchi\footnote{electronic address: takeuchi@vt.edu}
}
\affiliation{Institute for Particle Physics and Astrophysics,
Physics Department, Virginia Tech, Blacksburg, VA 24061}

\date{December 31, 2001; Revised: March 15, 2001; 2nd revision: September 24, 2009}

\begin{abstract}
We investigate the effect of the minimal length uncertainty relation, 
motivated by perturbative string theory, 
on the density of states in momentum space.
The relation is implemented through the modified commutation relation
$[\hat{x}_i,\hat{p}_j]=i\hbar[(1+\beta\hat{p}^2)\delta_{ij}
+ \beta'\hat{p}_i\hat{p}_j]$.
We point out that this relation, which is an example of an UV/IR relation, 
implies the finiteness of the cosmological constant.
While our result does not solve the cosmological constant problem,
it does shed new light on the relation between
this outstanding problem and UV/IR correspondence.
We also point out that the blackbody radiation spectrum
will be modified at higher frequencies, but the effect is
too small to be observed in the cosmic microwave background
spectrum.
\end{abstract}


\maketitle
\section{Introduction}

In this paper, we continue our investigation \cite{CMOT1}
of the consequences of the commutation relation
\begin{equation}
[ \hat{x}, \hat{p} ] = i\hbar (1 + \beta \hat{p}^2) \;,
\label{Eq:Com1}
\end{equation}
which leads to the minimal length uncertainty relation
\begin{equation}
\Delta x \ge \frac{\hbar}{2}
             \left( \frac{1}{\Delta p} + \beta\,\Delta p
             \right)\;.
\label{Uncert}
\end{equation}
As reviewed in Ref.~\cite{CMOT1}, Eq.~(\ref{Uncert}) has appeared in the 
context of perturbative string theory \cite{gross} where it is 
implicit in the fact that strings cannot probe distances 
below the string scale $\hbar\sqrt{\beta}$. 
It should be noted that the precise theoretical framework for such
a minimal length uncertainty relation is not understood in string theory. 
In particular, it is not clear whether Eq.~(\ref{Eq:Com1}) represents the 
correct quantum mechanical implementation of Eq.~(\ref{Uncert}).
Indeed, Kempf has shown that the commutation relation which implies the
existence of a minimal length is not unique \cite{Kempf:2001}.

Furthermore, Eq.~(\ref{Uncert}) does not seem to be universally valid.
For example, both in the realms of perturbative and
non-perturbative string theory (where distances shorter than
string scale can be probed by $D$-branes \cite{joep}), 
another type of uncertainty relation involving both spatial and time 
coordinates has been found to hold \cite{stu}. 
The distinction (and relation) between the minimal length uncertainty 
relation and the space-time uncertainty relation 
has been clearly emphasized by Yoneya \cite{yoneya}.

Notwithstanding these caveats, the minimum length uncertainty 
formula does exhibit the basic features of
UV/IR correspondence: when $\Delta p$ is large, 
$\Delta x$ is proportional to $\Delta p$, 
a fact which seems counterintuitive from the point of view of
local quantum field theory.
As is well known, this kind of UV/IR correspondence has been previously
encountered in various contexts: 
the $AdS/CFT$ correspondence \cite{adscft}, 
non-commutative field theory \cite{ncft},
and more recently in the attempts to understand quantum
gravity in asymptotically de Sitter spaces \cite{dscft,dsrg}.

Moreover, it has been argued by various authors \cite{banks} that the
UV/IR correspondence, described by Eq.~(\ref{Uncert}), is relevant for 
the understanding of the cosmological constant problem \cite{cosmoc}.  
Likewise, it has been suggested in the literature that some kind of 
UV/IR relation is necessary to understand observable
implications of short distance physics on inflationary
cosmology \cite{Kempf:2001,greene}.

In this paper we ask the question whether the cosmological
constant problem could be understood by utilizing
a concrete UV/IR relation, such as Eq.~(\ref{Eq:Com1}).
In particular, we study the implication of 
the commutation relation on the
effective density of states in the vacuum and consequently
on the cosmological constant problem. 
We point out that the commutation relation implies 
the finiteness of the cosmological constant and the modification 
of the blackbody radiation spectrum.
While we do not present a solution to the cosmological 
constant problem, our results offer a new perspective 
from which this outstanding problem may be addressed.

\section{The Classical Limit and the Liouville Theorem}

The observation we would like to make
is that the right hand side of Eq.~(\ref{Eq:Com1})
can be considered to define an `effective' value of $\hbar$ which 
is $p$-dependent.
This means that the size of the unit cell that each quantum state occupies 
in phase space can be thought of as being also 
$p$-dependent.
This will change the $p$-dependence of the density of states and
affect the calculation of the cosmological constant, the blackbody
radiation spectrum, etc. \cite{Lubo:2000yj}.
For this interpretation to make sense, we must first check that
any volume of phase space evolves in such a way that the number of
states inside does not change with time.
What we are looking for here is the analog of the Liouville theorem.
To place the discussion in a general context, we begin by extending
Eq.~(\ref{Eq:Com1}) to higher dimensions.

In $D$-dimensions, 
Eq.~(\ref{Eq:Com1}) is extended to the tensorial form \cite{Kempf:1995su} :
\begin{equation}
[ \hat{x}_i, \hat{p}_j ]
= i\hbar( \delta_{ij}
          + \beta \hat{p}^2 \delta_{ij}
          + \beta' \hat{p}_i \hat{p}_j
        )\;.
\label{Eq:Com2}
\end{equation}
If the components of the momentum $\hat{p}_i$ are assumed
to commute with each other, 
\begin{equation}
[ \hat{p}_i, \hat{p}_j ] = 0\;,
\label{Eq:Com3}
\end{equation}
then the commutation relations among the coordinates $\hat{x}_i$ 
are almost uniquely determined by the Jacobi Identity  
(up to possible extensions) as 
\begin{equation}
[ \hat{x}_i, \hat{x}_j ]
= i\hbar\,
  \frac{(2\beta-\beta') + (2\beta+\beta')\beta\hat{p}^2}
       { (1+\beta \hat{p}^2) }
  \left( \hat{p}_i \hat{x}_j - \hat{p}_j \hat{x}_i
  \right)\;.
\label{Eq:Com4}
\end{equation}

Let us take a look at what happens in the classical limit.
Recall that the quantum mechanical commutator corresponds to 
the Poisson bracket in classical mechanics via
\begin{equation}
\frac{1}{i\hbar} [ \hat{A}, \hat{B} ] \Longrightarrow \{A,B\}\;.
\end{equation}
So the classical limits of Eqs.~(\ref{Eq:Com2})--(\ref{Eq:Com4}) read
\begin{eqnarray}
\{x_i,p_j\} & = & ( 1 + \beta p^2 ) \delta_{ij} + \beta' p_i p_j \;,\cr
\{p_i,p_j\} & = & 0 \;,\cr
\{x_i,x_j\} & = & \frac{(2\beta-\beta') + (2\beta+\beta')\beta p^2}
                       { (1+\beta p^2) }
                  \left( p_i x_j - p_j x_i \right)\;.
\label{Eq:Poi5}
\end{eqnarray}
The time evolutions of the coordinates and momenta are governed by
\begin{eqnarray}
\dot{x}_i & = & \{x_i,H\} 
\;=\; \phantom{-}\{x_i,p_j\}\,\frac{\partial H}{\partial p_j} 
    + \{x_i,x_j\}\,\frac{\partial H}{\partial x_j} \;,\cr
\dot{p}_i & = & \{p_i,H\}
\;=\; -\{x_j,p_i\}\,\frac{\partial H}{\partial x_j} \;.
\end{eqnarray}
The analog of the Liouville theorem in this case states that the
weighted phase space volume 
\begin{equation}
\frac{ d^D\mathbf{x}\,d^D\mathbf{p} }
     { \Bigl[ 1+\beta p^2 \Bigr]^{D-1}
       \Bigl[ 1+(\beta+\beta') p^2 \Bigr] 
     } 
\label{INVARIANT}
\end{equation}
is invariant under time evolution.
To see this, consider an infinitesimal time interval $\delta t$. 
The evolution of the coordinates and momenta during $\delta t$ are
\begin{eqnarray}
x_i' & = & x_i + \delta x_i \;, \cr
p_i' & = & p_i + \delta p_i \;,
\end{eqnarray}
with
\begin{eqnarray}
\delta x_i 
& = & \left[ \{x_i,p_j\}\,\frac{\partial H}{\partial p_j} 
           + \{x_i,x_j\}\,\frac{\partial H}{\partial x_j}
      \right]\delta t \;, \cr
\delta p_i
& = & \left[ -\{x_j,p_i\}\,\frac{\partial H}{\partial x_j}
      \right]\delta t \;.
\end{eqnarray}
An infinitesimal phase space volume after this infinitesimal evolution is
\begin{equation}
d^D\mathbf{x}'\,d^D\mathbf{p}'
= \left| \dfrac{\partial(x'_1,\cdots,x'_D,p'_1,\cdots,p'_D)}
               {\partial(x_1, \cdots,x_D, p_1, \cdots,p_D)}
  \right|
d^D\mathbf{x}\,d^D\mathbf{p} \;.
\end{equation}
Since
\begin{equation}
\begin{array}{ll}
\dfrac{\partial x'_i}{\partial x_j}
= \delta_{ij} + \dfrac{\partial \delta x_i}{\partial x_j}\;,
&\qquad
\dfrac{\partial x'_i}{\partial p_j}
= \dfrac{\partial \delta x_i}{\partial p_j} \;, \cr
\dfrac{\partial p'_i}{\partial x_j}
= \dfrac{\partial \delta p_i}{\partial x_j}\;,
&\qquad
\dfrac{\partial p'_i}{\partial p_j}
= \delta_{ij} + \dfrac{\partial \delta p_i}{\partial p_j} \;,
\end{array}
\end{equation}
the Jacobian to first order in $\delta t$ is
\begin{equation}
\left| \dfrac{\partial(x'_1,\cdots,x'_D,p'_1,\cdots,p'_D)}
               {\partial(x_1, \cdots,x_D, p_1, \cdots,p_D)}
\right| =
1 + \left( \frac{\partial\delta x_i}{\partial x_i}
         + \frac{\partial\delta p_i}{\partial p_i}
    \right)
  + \cdots \;.
\end{equation}
We find
\begin{eqnarray}
\lefteqn{
\left(\frac{\partial\delta x_i}{\partial x_i}
    + \frac{\partial\delta p_i}{\partial p_i}
\right)\frac{1}{\delta t} } \cr
& = & \frac{\partial}{\partial x_i}
      \left[ \{x_i,p_j\}\,\frac{\partial H}{\partial p_j} 
           + \{x_i,x_j\}\,\frac{\partial H}{\partial x_j}
      \right]
    - \frac{\partial}{\partial p_i}
      \left[ \{x_j,p_i\}\,\frac{\partial H}{\partial x_j}
      \right] \cr
& = & \left[ \frac{\partial}{\partial x_i}\{x_i,p_j\} \right]
      \frac{\partial H}{\partial p_j}
    + \{x_i,p_j\}\frac{\partial^2 H}{\partial x_i \partial p_j}
    + \left[ \frac{\partial}{\partial x_i}\{x_i,x_j\} \right]
      \frac{\partial H}{\partial x_j} \cr
& & + \{x_i,x_j\}\frac{\partial^2 H}{\partial x_i \partial x_j}
    - \left[ \frac{\partial}{\partial p_i}\{x_j,p_i\} \right]
      \frac{\partial H}{\partial x_j}
    - \{x_j,p_i\}\frac{\partial^2 H}{\partial p_j \partial x_i} \cr
& = & \left[ \frac{\partial}{\partial x_i}\{x_i,x_j\} \right]
      \frac{\partial H}{\partial x_j}
    - \left[ \frac{\partial}{\partial p_i}\{x_j,p_i\} \right]
      \frac{\partial H}{\partial x_j} \cr
& = & \Biggl[ -\frac{ (2\beta-\beta') + (2\beta+\beta')\beta p^2 }
            { (1+\beta p^2) } (D-1)\,p_j
      \Biggr] \frac{\partial H}{\partial x_j}
     -\Biggl[ \{2\beta + (D+1)\beta'\}\,p_j 
      \Biggr]\frac{\partial H}{\partial x_j} \cr
& = & -\left[ 2(D-1)\beta
               \left( \frac{ 1 + (\beta+\beta') p^2 }
                           { 1 + \beta p^2 }
               \right)
             + 2(\beta + \beta')
       \right] p_j \frac{\partial H}{\partial x_j} \;.
\end{eqnarray}
Therefore, to first order in $\delta t$
\begin{equation}
d^D\mathbf{x}'\,d^D\mathbf{p}'
= d^D\mathbf{x}\,d^D\mathbf{p} 
  \left[ 1 
     - \left\{ 2(D-1)\beta\left( \frac{ 1 + (\beta+\beta') p^2 }
                                      { 1 + \beta p^2 }
                          \right)
             + 2(\beta + \beta')
       \right\} p_j\frac{\partial H}{\partial x_j}\delta t
  \right] \;.
\label{DXDP}
\end{equation}
On the other hand,
\begin{eqnarray}
1 + \beta {p'}^2
& = & 1 + \beta (p_i + \delta p_i)^2 \cr
& = & 1 + \beta\left( p^2 + 2p_i \delta p_i + \cdots \right) \cr
& = & 1 + \beta\left( p^2 
                  - 2p_i\{x_i,p_j\}\frac{\partial H}{\partial x_j}\delta t
                  + \cdots
                \right) \cr
& = & 1 + \beta\left( p^2
                - 2\left[ 1+(\beta+\beta') p^2
                   \right]
                    p_j\frac{\partial H}{\partial x_j}\delta t
                + \cdots  
               \right)  \cr
& = & (1+\beta p^2)
       - 2\beta\left[ 1+(\beta+\beta') p^2
               \right]
       p_j\frac{\partial H}{\partial x_j}\delta t + \cdots \cr
& = & (1+\beta p^2)
      \left[ 1 - 2\beta
                  \left( \frac{1+(\beta+\beta') p^2}{1+\beta p^2}
                  \right)
             p_j\frac{\partial H}{\partial x_j}\delta t + \cdots
      \right]\;,
\end{eqnarray}
and
\begin{eqnarray}
1 + (\beta+\beta') {p'}^2
& = & 1 + (\beta+\beta') (p_i + \delta p_i)^2 \cr
& = & 1 + (\beta+\beta')\left( p^2 + 2p_i \delta p_i + \cdots \right) \cr
& = & 1 + (\beta+\beta')
          \left( p^2 
               - 2p_i\{x_i,p_j\}\frac{\partial H}{\partial x_j}\delta t
               + \cdots
          \right) \cr
& = & 1 + (\beta+\beta')
          \left( p^2
               - 2\left[ 1+(\beta+\beta') p^2
                  \right] 
                  p_j\frac{\partial H}{\partial x_j}\delta t
               + \cdots 
          \right) \cr 
& = & \left[ 1+(\beta+\beta') p^2 \right]
     - 2(\beta+\beta') \left[ 1+(\beta+\beta') p^2 \right]
       p_j\frac{\partial H}{\partial x_j}\delta t + \cdots \cr
& = & [1+(\beta+\beta') p^2]
      \left[ 1 - 2(\beta+\beta') 
                 p_j\frac{\partial H}{\partial x_j}\delta t + \cdots
      \right]\;.
\end{eqnarray}
Therefore, to first order in $\delta t$
\begin{eqnarray}
\lefteqn{
\Bigl[ 1 + \beta{p'}^2 \Bigr]^{-D+1}
\Bigl[ 1 + (\beta+\beta') {p'}^2 \Bigr]^{-1}
}\cr
& = & 
\Bigl[ 1 + \beta p^2 \Bigr]^{-D+1}
\Bigl[ 1 + (\beta+\beta') p^2 \Bigr]^{-1} \cr
& & \times
\left[ 1 + 
       \left\{
           2(D-1)\beta\left( \frac{ 1+(\beta+\beta')p^2 }
                                  { 1+\beta p^2 }
                      \right)
         + 2(\beta+\beta')
       \right\} p_j\frac{\partial H}{\partial x_j}\delta t
\right] \;.
\label{WEIGHT}
\end{eqnarray}
From Eqs.~(\ref{DXDP}) and (\ref{WEIGHT}), we deduce that
the weighted phase space volume Eq.~(\ref{INVARIANT}) is
invariant. 
Note that in addition to the non-canonical Poisson brackets between the
coordinates and momenta, those among the coordinates themselves 
(and thus the non--commutative geometry of the problem) is crucial 
in arriving at this result.
When $\beta'=0$, Eq.~(\ref{INVARIANT}) simplifies to
\begin{equation}
\frac{ d^D\mathbf{x}\,d^D\mathbf{p} }
     { (1+\beta p^2)^D } \,. 
\label{INVARIANT2}
\end{equation}
%


As a concrete example, consider the $1D$ harmonic oscillator
with the Hamiltonian
\begin{equation}
H = \frac{p^2}{2\mu} + \frac{1}{2}\mu\omega^2 x^2 \;.
\end{equation}
The equations of motion are
\begin{eqnarray}
\dot{x} 
& = & \{ x, H \}
\;=\; \frac{1}{\mu} (1+\beta p^2) p \;, \cr
\dot{p}
& = & \{ p, H \}
\;=\; -\mu\omega^2(1+\beta p^2) x \;.
\end{eqnarray}
These equations can be solved to yield
\begin{eqnarray}
x(t) & = & x_{\max}\,\sqrt{1+\varepsilon} \, 
\dfrac{ \sin(\sqrt{1+\varepsilon}\,\omega t) }
      { \sqrt{ 1 
             + \varepsilon\,\sin^2(\sqrt{1+\varepsilon}\,\omega t)
             }
      } \;,\cr
p(t) & = & p_{\max}\, 
\dfrac{ \cos(\sqrt{1+\varepsilon}\,\omega t) }
      { \sqrt{ 1
             + \varepsilon\,\sin^2(\sqrt{1+\varepsilon}\,\omega t)
             }
      } \;,
\end{eqnarray}
where
\begin{equation}
\varepsilon = 2\mu E \beta \;, \qquad
x_{\max} = \sqrt{ \frac{2E}{\mu\omega^2} } \;,\qquad
p_{\max} = \sqrt{ 2\mu E } \;.
\end{equation}
Note that the period of oscillation, $T$, is now energy (and thus amplitude)
dependent:
\begin{equation}
T = \frac{2\pi}{\omega\sqrt{1+\varepsilon}}\;.
\end{equation}
Now, consider the infinitesimal phase space volume sandwiched between the
equal--energy contours $E$ and $E+dE$, 
and the equal--time contours $t$ and $t+dt$. 
It is straightforward to show that
\begin{equation}
dE\,dt = \frac{dx\,dp}{1+\beta p^2}\;.
\end{equation}
The left--hand side of this equation is time--independent by 
definition, so the right--hand side must be also.

Finally, note that the semi--classical quantization
($\hbar \rightarrow 0$ limit)
of the harmonic oscillator is consistent with the
full quantum mechanical result derived in our
previous paper, Ref.~\cite{CMOT1}.

\section{Density of States}

From this point on, we will only consider the $\beta'=0$ case
for the sake of simplicity.

\begin{figure}[ht]
\begin{center}
\unitlength=1cm
\begin{picture}(12,7)(0,0)
\put(10.5,0.35){$\log_{10}(\sqrt{\beta} p)$}
\includegraphics{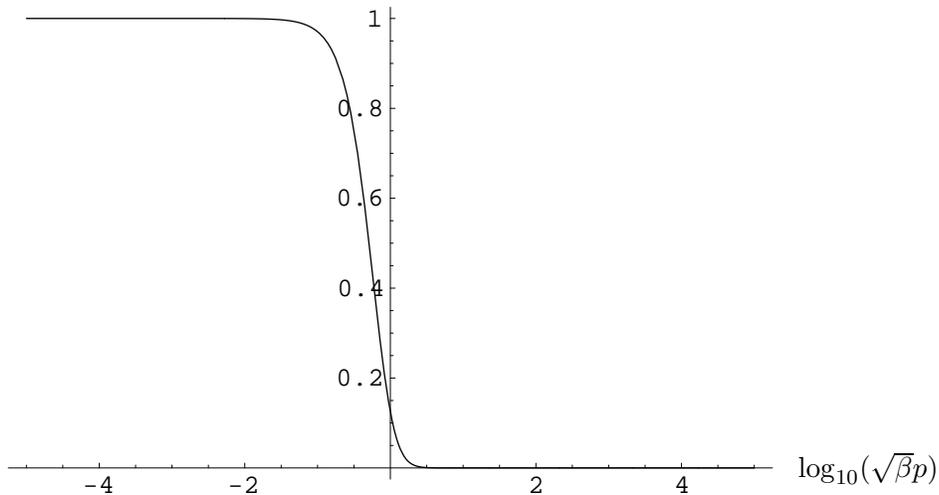}
\end{picture}
\caption{The behavior of the weight factor
$[1+(\sqrt{\beta} p)^2]^{-3}$.} 
\label{DAMP}
\end{center}
\end{figure}

Integrating over the coordinates, 
the invariant phase space volume Eq.~(\ref{INVARIANT2}) becomes
\begin{equation}
\frac{V\,d^D\mathbf{p}}{(1+\beta p^2)^D} \;, 
\end{equation}
where $V$ is the coordinate space volume.
This implies that upon quantization, the number of quantum states
per momentum space volume should be assumed to be
\begin{equation}
\frac{V}{(2\pi\hbar)^D}\frac{d^D\mathbf{p}}{(1+\beta p^2)^D} \;.
\label{DENSITY}
\end{equation}
Eq.~(\ref{DENSITY}) indicates that the density of states in momentum space 
must be modified by the extra factor of $(1+\beta p^2)^{-D}$.
This factor effectively cuts off the integral beyond
$p = 1/\sqrt{\beta}$.
Indeed, in $3D$ the weight factor is
\begin{equation}
\frac{1}{(1+\beta p^2)^3}\;,
\label{3Dweight}
\end{equation}
the plot of which is shown in Fig.~\ref{DAMP}.
We look at the consequence of this modification in the calculation of the
cosmological constant and the blackbody radiation spectrum in the following.

\subsection{The Cosmological Constant}

The cosmological constant is obtained by summing over the zero--point
fluctuation energies of harmonic oscillators, each of which corresponds
to a particular particle momentum state \cite{cosmoc}. 
If we assume that the zero--point energy of each oscillator is of the
usual, canonical, form
\begin{equation}
\frac{1}{2}\hbar\omega = \frac{1}{2}\sqrt{p^2 + m^2} \;,
\end{equation}
then, the sum over all momentum states per unit volume is
(up to some prefactors)
\begin{eqnarray}
\Lambda(m) 
& = & \int \frac{d^3\mathbf{p}}{(1+\beta p^2)^3}
      \left[ \frac{1}{2} \sqrt{p^2 + m^2} \right] \cr
& = & 2\pi\int_0^\infty \frac{p^2 dp}{(1+\beta p^2)^3}\,\sqrt{p^2 + m^2} \cr
& = & \frac{\pi}{2\beta^2}\,f(\beta m^2)\;,
\label{CosmoConstant}
\end{eqnarray}
where
%
%
\begin{equation}
f(x) =
\left\{ \begin{array}{ll}
        1 + \dfrac{x}{2(1-x)}
          + \dfrac{x^2}{4(1-x)^{3/2}}
            \ln\left(\dfrac{ 1 - \sqrt{ 1-x } }
                           { 1 + \sqrt{ 1-x } }
               \right)
        &\qquad (x \le 1)\;, \cr
        1 - \dfrac{x}{2(x-1)}
          + \dfrac{x^2}{2(x-1)^{3/2}}
            \tan^{-1}\sqrt{ x-1 }
        &\qquad (x \ge 1)\;. \cr
        \end{array}
\right. 
\end{equation}
$f(x)$ is a monotonically increasing function which behaves 
asymptotically as
\begin{equation}
f(x)\sim \sqrt{x}\;,
\end{equation}
as can be gleaned from Eq.~(\ref{CosmoConstant}).
In the region $0\le x\le 1$, it is well approximated by
\begin{equation}
f(x)\approx (1+x)^{0.42}\;.
\end{equation}
In the massless case we obtain
\begin{equation}
\Lambda(0) = \frac{\pi}{2\beta^2} \;.
\end{equation}
As expected, due to the strong suppression of the density of states
at high momenta, the cosmological constant is
rendered finite with $1/\sqrt{\beta}$ acting effectively as
the UV cutoff.  This result is in strong contrast to
conventional calculations where the UV cutoff is an arbitrary
scale which must be introduced by hand, and where one must assume 
that the physics beyond the cutoff does not contribute.

Unfortunately, since $1/\sqrt{\beta}$ is the string mass scale,
which we expect to be of the order of the Planck mass $M_p$, 
this does not solve the cosmological constant problem.
This is true even if the Planck mass $M_p$ were as low as a TeV as 
suggested in models with large extra dimensions \cite{ADD}.

\subsection{The Blackbody Radiation Spectrum}

Taking into account the weight factor Eq.~(\ref{3Dweight}), 
the average energy in the EM field per unit volume at temperature $T$ is
\begin{eqnarray}
\bar{E}
& = & 2\int \frac{d^3\mathrm{k}}{(2\pi)^3[\,1+\beta(\hbar k)^2\,]^3}\,
      \dfrac{\hbar k c}{e^{\hbar k c/k_B T} -1 } \cr
& = & \frac{8\pi}{c^3}\int_0^\infty d\nu\,
      \frac{1}{[\,1+\beta(h\nu/c)^2\,]^3}
      \left(\frac{h\nu^3}{e^{h\nu/k_B T} - 1}\right) \cr
& \equiv & \int_0^\infty d\nu\,u_\beta(\nu,T)\;.
\end{eqnarray}
We see that the blackbody radiation spectrum is damped at high
frequencies close to the cutoff scale:
\begin{equation}
u_\beta(\nu,T)
= \frac{1}{[1+(\nu/\nu_\beta)^2]^3}\,u_0(\nu,T) \;,\qquad
\nu_\beta \equiv \frac{c}{h\sqrt{\beta}}\;.
\end{equation}
Here
\begin{equation}
u_0(\nu,T)
= \frac{8\pi h \nu^3}{c^3}\frac{1}{e^{h\nu/k_B T}-1} \;,
\end{equation}
is the regular spectral function.

\begin{figure}[ht]
\begin{center}
\unitlength=1cm
\begin{picture}(16,4.5)(0,0)
\put(3.4,-0.2){$\nu/\nu_\beta$}
\put(10.6,-0.2){$\nu/\nu_\beta$}
\put(5.2,4){$T=T_\beta$}
\put(12,4){$T=0.1\,T_\beta$}
\resizebox{7cm}{!}{\includegraphics[bb=88 4 376 182]{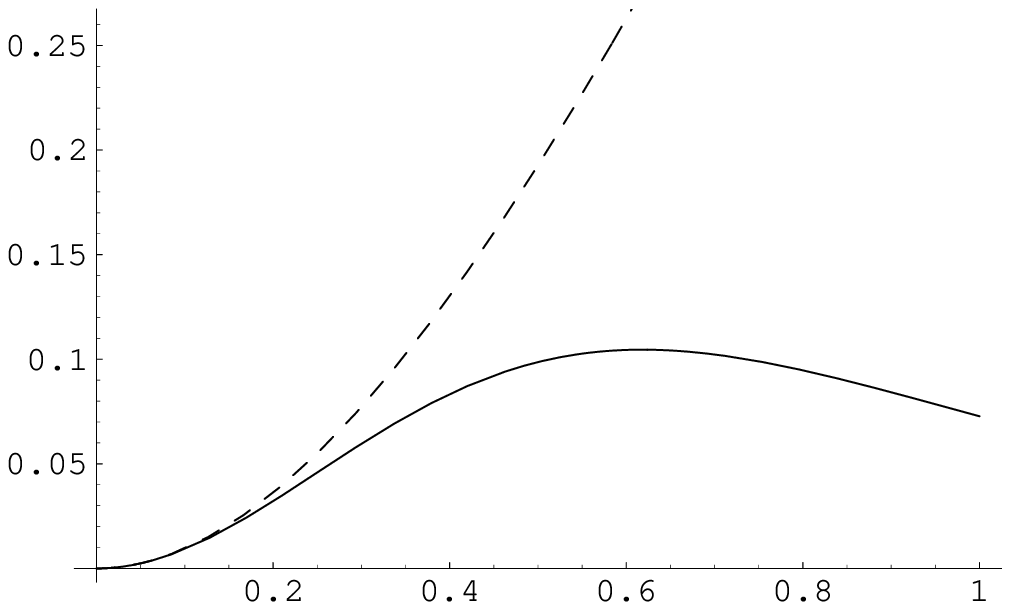}}
\resizebox{7cm}{!}{\includegraphics[bb=88 4 376 182]{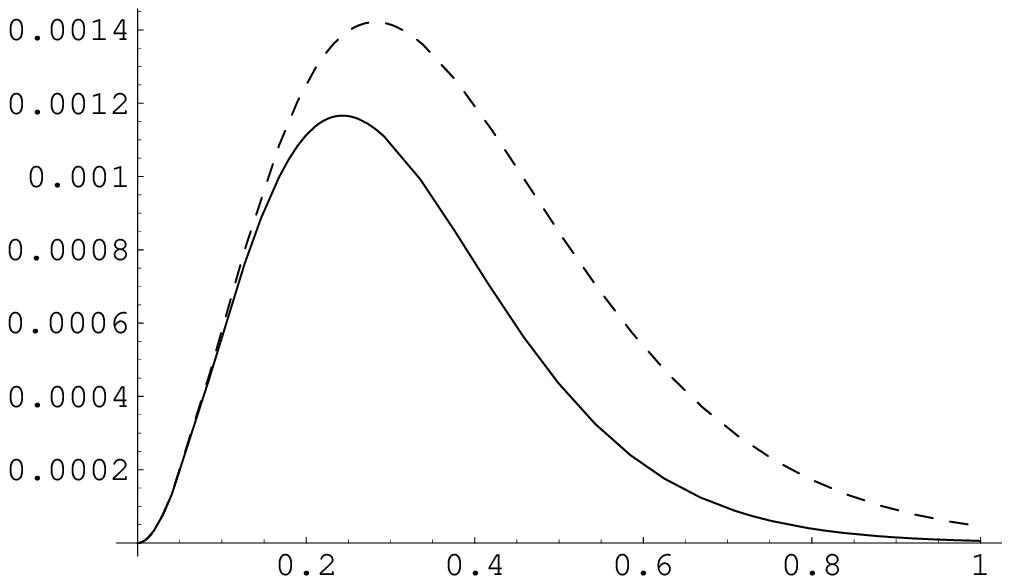}}
\end{picture}
\caption{The shape of the blackbody radiation spectrum 
with (solid line) and without (dashed line) the damping factor
$[1+(\nu/\nu_\beta)^2]^{-3}$ at temperatures
$T=T_\beta$ (left) and $T=0.1\,T_\beta$ (right).} 
\label{BB1}
\end{center}
\end{figure}

\begin{figure}[ht]
\begin{center}
\unitlength=1cm
\begin{picture}(16,4.5)(0,0)
\put(3.4,-0.2){$\nu/\nu_\beta$}
\put(10.8,-0.2){$\nu/\nu_\beta$}
\put(4.6,4.1){$T=0.01\,T_\beta$}
\put(11.8,4.1){$T=0.01\,T_\beta$}
\resizebox{7cm}{!}{\includegraphics[bb=88 4 376 182]{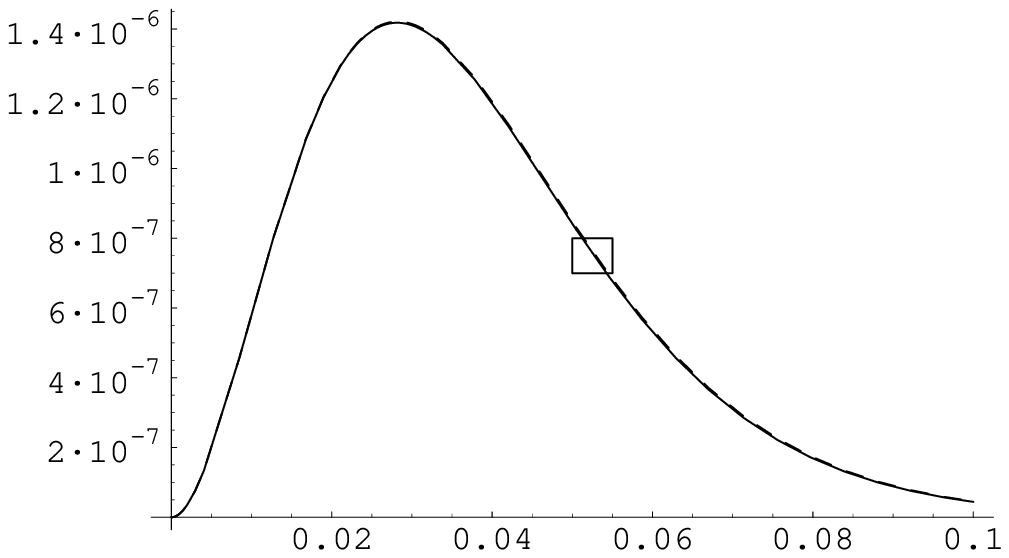}}
\resizebox{7cm}{!}{\includegraphics[bb=88 4 376 182]{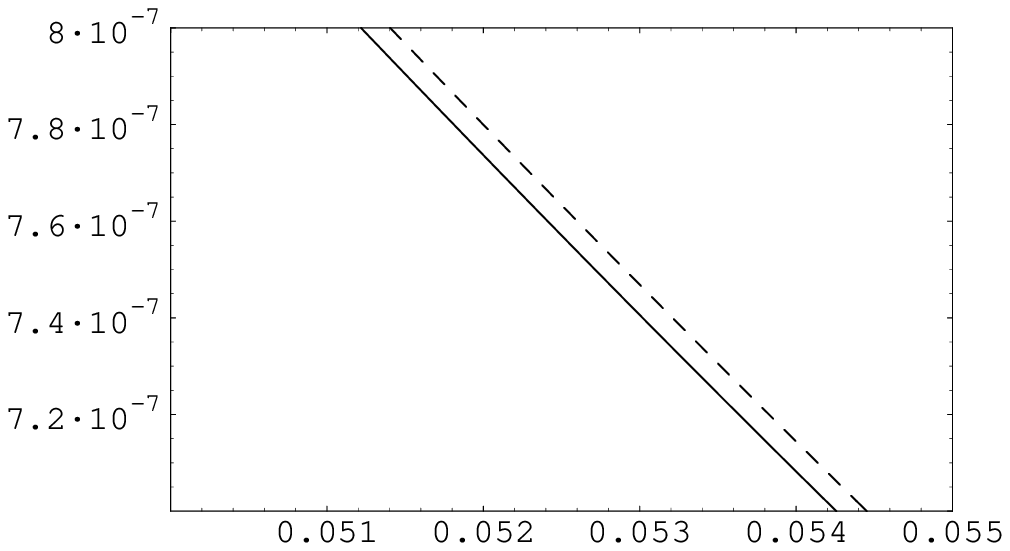}}
\end{picture}
\caption{The shape of the blackbody radiation spectrum
with (solid line) and without (dashed line) the damping factor
$[1+(\nu/\nu_\beta)^2]^{-3}$ at temperature
$T=0.01\,T_\beta$.  The graph on the right is the blow-up of the
region inside the box on the left.} 
\label{BB2}
\end{center}
\end{figure}

To see the effect of this damping on the shape of the spectral function,
we plot the functions 
\begin{eqnarray}
f_0(\nu,T) 
& \equiv & \dfrac{ (\nu/\nu_\beta)^3 }
                 { e^{(\nu/\nu_\beta)(T_\beta/T)} - 1 } \;,\cr
f_\beta(\nu,T)
& \equiv & \dfrac{ 1 }{ [1+(\nu/\nu_\beta)^2]^3 }\,f_0(\nu,T) \;,
\end{eqnarray}
for several values of the temperature 
$T$ in Figs.~\ref{BB1} and \ref{BB2}.  
The temperature $T_\beta$ in the definition of $f_0(\nu,T)$ is 
defined as
\begin{equation}
T_\beta \equiv \frac{c}{k_B \sqrt{\beta}} \;.
\end{equation}
As is evident from the figures, the distortion to 
the blackbody radiation is undetectable unless the temperature is 
within a few orders of magnitude below $T_\beta$.
Given that the decoupling temperature is on the order of an MeV \cite{SMOOT},
we do not expect the spectrum of the Cosmic Microwave Background (CMB) 
to be affected in any observable way, even if $1/\sqrt{\beta}$ 
were as small as a TeV \cite{ADD}.

\section{Discussion}

We have shown that Eq.~(\ref{Eq:Com1}) and its higher dimensional
extensions imply that the density of states is naturally suppressed
in the ultraviolet.  This suppression renders the cosmological
constant finite, without affecting the blackbody radiation spectrum
at observable temperatures.  
The cosmological constant is still too large,
however, since it is proportional to $1/\beta^2 \sim M_p^4$.
This scaling is to be expected since $1/\sqrt{\beta}$ is the only
scale present in the problem.

Nevertheless, we believe that
our result is a major improvement over previous results
and points to a new direction from which we may approach the cosmological 
constant problem.  The reason why conventional calculations were
obtaining an infinite result can be identified as a case of over-counting
of states. 
If string theory does indeed lead to Eq.~(\ref{Eq:Com1}) and the
resulting suppression of the density of states at high momenta,
the number of states in the UV is not as numerous as conventional
calculations assume.

The smallness of the cosmological constant can then be interpreted as a
sign that we are still over-counting the number of states in our calculation 
and the density of states should be further suppressed.   
Whatever the method of suppression,
this would necessarily entail the introduction of
an additional characteristic scale, $M_c$, 
other than $1/\sqrt{\beta} = M_p$, in which case the dependence of the
cosmological constant on the scales can be expressed as
$\Lambda \sim M_p^4 F(M_c/M_p)$ with $F(1)=1$.

One method we could use would be to modify the right hand side of 
Eq.~(\ref{Eq:Com1}) to further suppress the higher momentum states
\footnote{Depending on the modification, the cosmological constant
may be rendered divergent.  For instance, if we adopt the commutation relations
used by Kempf in Ref.~\cite{Kempf:2001}, the cosmological constant
will diverge logarithmically.}.
However, that would not work since all that would do would be to
change the `effective' value of the UV cutoff from $M_p$ to $M_c$.
($F\sim (M_c/M_p)^4$.)
So the additional suppression must occur in a more subtle fashion.

Banks \cite{banks} has argued that $F\sim (M_c/M_p)^8$ with 
$M_c \sim 1 TeV$ and $M_p \sim 10^{19} GeV$ would reproduce the correct 
value of the cosmological constant.  But the exact mechanism
which would lead to such a form for $F$ remains elusive.
(See also Ref.~\cite{per} for a related discussion.)

At this point, we note that if string theory is the correct theory
of gravity, modifying the commutation relations alone would not
properly take into account all of its potential effects.
It is possible that the holographic principle \cite{holography} could be
of help here.   Consideration of holography in a cosmological background 
might naturally provide another scale other than $1/\sqrt{\beta}$, namely 
the size of the cosmological horizon, the Hubble radius $H$, related to 
the cosmological constant as $H^2 \sim 1/\Lambda$. 
It is conceivable that due to the correct implementation
of the holographic principle in a cosmological situation, 
the number of fundamental degrees of freedom contributing to the vacuum 
energy is determined by the density of states above some very large momentum 
(which by the UV/IR correspondence (\ref{Uncert}) would be related to the
degrees of freedom at distances of the order of the cosmological horizon).
If indeed the density of states is strongly suppressed at high momenta, 
as argued in this paper, 
then the effective number of degrees of freedom contributing to the 
vacuum energy density would be very small. 
While these considerations are highly speculative, they seem to point 
to a new promising way to approach the cosmological constant problem.

\newpage
\acknowledgments

We would like to thank Vijay Balasubramanian, 
Per Berglund, Will Loinaz, Asad Naqvi, 
Koenraad Schalm, Gary Shiu, Joseph Slawny,
and Matthew Strassler for helpful discussions.
This research is supported in part by a grant from the US 
Department of Energy, DE--FG05--92ER40709.


\end{document}